\documentclass[aps,pra,twocolumn,amsmath,amssymb]{revtex4-1}
\usepackage{mathrsfs}
\usepackage{graphicx}
\usepackage{dcolumn}
\usepackage{bm}
\usepackage{amsmath}
\usepackage{amsfonts}

\begin{document}

\title{Spontaneous Spin Textures in Dipolar Spinor Condensates:  A Dirac String Gas Approach}
\author{Biao Lian}
\affiliation{Department of Physics, McCullough Building, Stanford University, Stanford, California 94305-4045, USA}
\date{\today}

\begin{abstract}
We study the spontaneous spin textures induced by magnetic dipole-dipole interaction in ferromagnetic spinor condensates under various trap geometries. At the mean-field level, we show the system is dual to a Dirac string gas with a negative string tension in which the ground state spin texture can be easily determined. We find that three-dimensional condensates prefer a meron-like vortex texture, quasi one-dimensional condensates prefer the axially polarized flare texture, while condensates in quasi two dimensions exhibit either a meron texture or an in-plane polarized texture.

\end{abstract}
\maketitle

\section{Introduction}
The magnetic dipole-dipole interaction (DDI) is known to play a significant role in determining the long range behaviors of ultracold gases of spinful atoms, especially those with a large spin \cite{Stuhler2005,Lu2011,Aikawa2012}. In the ferromagnetic spinor condensates of spinful bosons with nonzero spin expectation values $\langle \mathbf{F}\rangle$, the DDI usually leads to rich spin textures that strongly depend on the trap geometry and the initial state preparation \cite{Vengalattore2008,Vengalattore2010,Kronjager2010,Eto2014}. Meanwhile, the existence of spontaneous spin textures in dipolar spinor condensates has been verified by numerical simulations based on the mean-field Gross-Pitaevskii equation \cite{Yi2006,Zhang2010,Takahashi2007,Kawaguchi2010,Huhtamaki2010}. However, due to the complicated form of DDI, there still lacks an explicit and efficient theoretical way to understand the spin textures.

In this letter, we develop an semi-analytical approach to the problem by introducing a duality at the mean-field level from a strong dipolar spinor condensate to a weakly interacting Dirac string gas with a negative string tension.
%The duality is a strong-weak duality in the sense that a condensate with a strong magnetic dipole-dipole interaction is dual to a weakly interacting Dirac string gas. Since the string tension is negative, the space is fulfilled with Dirac strings.
We find that the ground state is reached by forming as many closed strings with small enough curvatures as possible. Based on this principle, we are able to determine the spin textures in various trap geometries used in cold atom experiments.
%study the spontaneous spin textures in several trap geometries commonly adopted in the experiments.
We show the ground state spin texture of a large three-dimensional (3D) spherical condensate is a meron-like vortex texture. Further, we show the axially polarized flare texture \cite{Takahashi2007} is favored in quasi one-dimensional (1D) condensates, while
in quasi two-dimensional (2D) pancake traps there is a phase transition between an in-plane polarized texture and a meron texture driven by the pancake radius. In particular, multiple merons can be seen to arise in quasi 2D traps when the aspect ratio is comparatively large, in agreement with the numerical results in Ref. \cite{Zhang2010}. The Dirac string gas picture thus offers a greatly simplified way to understand the spin textures in dipolar condensates.

\section{Theoretical Formulation}\label{theory}
We begin with the description of the ferromagnetic spinor condensates at low energies. In general, a spinor condensate of integral spin $F$ atoms is characterized by a spinor order parameter of $2F+1$ components $\Psi_m(\mathbf{r})$ ($m=-F, -F+1,\cdots, F$), which represents the coherent amplitude of annihilating a boson in the spin state $|F,m\rangle$ at position $\mathbf{r}$ \cite{Ho1998,Ohmi1998,Kawaguchi2011,Lian2012}. At the mean field level, the number of atoms per unit volume in the condensate is $n(\mathbf{r})=\sum_m\Psi_m^*(\mathbf{r})\Psi_m(\mathbf{r})=\Psi^\dag(\mathbf{r})\Psi(\mathbf{r})$, while the expectation of the local spin is $\langle \mathbf{F}(\mathbf{r})\rangle=\varphi^\dag(\mathbf{r})\mathbf{F}\varphi(\mathbf{r})$, where $\varphi(\mathbf{r})=\Psi(\mathbf{r})/\sqrt{n(\mathbf{r})}$ is the normalized spinor order parameter, and $\mathbf{F}$ is the spin matrix of the spin $F$ representation. In a square well trap potential or in the central region of a harmonic trap within a size $R\ll n(\mathbf{r})/|\nabla n(\mathbf{r})|$, we can approximate the particle number density $n(\mathbf{r})=n_0$ as a constant. Up to a U$(1)\times$SU$(2)$ transformation, the normalized spinor order parameter $\varphi(\mathbf{r})$ is determined by the local interactions, which are characterized by $F+1$ scattering lengths $a_{2J}$ between two atoms with total spin $2J$ ($0\le J\le F$) \cite{Lian2012}. A spinor condensate is said to be ferromagnetic if $\varphi(\mathbf{r})=(0,\cdots,0,1,0,\cdots,0,0)$ in a certain basis, where the only nonzero component is $m=F_{0}>0$.

At low energies, the long wave length fluctuation of the order parameter gives rise to multiple gapless Goldstone modes. In a ferromagnetic spinor condensate, there is exactly one quadratic dispersion mode corresponding to the fluctuation of the local spin \cite{Ho1998,Ohmi1998,Kawaguchi2011,Lian2012}. In the presence of the magnetic dipole-dipole interaction, the effective low energy spin Hamiltonian of a ferromagnetic spinor condensate can be written as $H=H_0+H_D$, where
\begin{equation}\label{HMF}
\begin{split}
&H_0=\frac{\alpha}{M}\int d^3\mathbf{r}(\nabla\bm{\mathcal{F}}(\mathbf{r}))^2\ ,\\
&H_D=\frac{\lambda}{2}\int d^3\mathbf{r}_1d^3\mathbf{r}_2\frac{\bm{\mathcal{F}}_1\cdot \bm{\mathcal{F}}_2-3(\bm{\mathcal{F}}_1\cdot \hat{\mathbf{r}}_{12})(\bm{\mathcal{F}}_2\cdot \hat{\mathbf{r}}_{12})}{4\pi r_{12}^3}
\end{split}
\end{equation}
are the kinetic energy of the quadratic Goldstone mode and the DDI energy, respectively (see Appendix \ref{MFdescription}). We have used here the normalized local spin field $\bm{\mathcal{F}}(\mathbf{r})=\langle \mathbf{F}(\mathbf{r})\rangle/F_0$ that satisfies $|\bm{\mathcal{F}}(\mathbf{r})|=1$ for later convenience. $M$ denotes the particle mass, $\alpha$ is defined by $\alpha=n_0\hbar^2[F(F+1)-F_0^2]/4$, and the interaction parameter $\lambda$ is given by $\lambda=\mu_0 (g_F\mu_B)^2n_0^2F_0^2$, where $g_F$ is the Land\'e factor and $\mu_B$ is the Bohr magneton. In writing the $H_D$ term, we have used the abbreviations $\bm{\mathcal{F}}_i=\bm{\mathcal{F}}(\mathbf{r}_i)$ and $\mathbf{r}_{12}=r_{12}\hat{\mathbf{r}}_{12}=\mathbf{r}_1-\mathbf{r}_2$, where $\hat{\mathbf{r}}_{12}$ is the corresponding unit vector.

\begin{figure}
\includegraphics[width=3in]{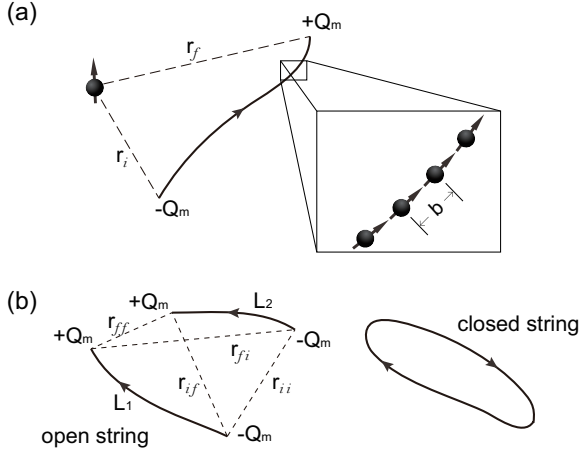}
\caption{(a) A single column chain of head-to-tail spins is equivalent to a Dirac string with two monopoles $\pm Q_m$ at the ends. (b) Under the dipolar interaction of spins, two open strings only have an interaction between the monopoles, while closed strings are free. \label{string}}
\end{figure}

It is difficult to infer the low energy spin texture configurations directly from the perspective of local spins, since the spin-spin interaction in $H_D$ is highly anisotropic. However, it is possible to understand $H_D$ more naturally in terms of Dirac strings. Consider a chain of atoms that are arranged uniformly along a curve as is shown in Fig. \ref{string}(a), with their spins aligned head-to-tail and parallel to the tangent of the curve. Since each atom carries a magnetic dipole moment $\mathbf{m}_F=g_F\mu_B\langle\mathbf{F}\rangle$, the curve can be exactly viewed as a Dirac string, with a positive magnetic monopole at one end and a negative one at the other. If we parameterize the curve as $\mathbf{r}(l)$ with an affine parameter $l$, and require $|d\mathbf{r}/dl|=1$, the local spin field on the curve is then $\bm{\mathcal{F}}=d\mathbf{r}/dl$. We can easily calculate the dipolar interaction between such a Dirac string and a spin magnetic moment $\mathbf{m}_F$ not belonging to the string:
\begin{equation}\label{EmF}
\begin{split}
&E_{\mathbf{m}_F}=\mu_0Q_m\int dl\frac{r^2 \mathbf{m}_F\cdot\frac{d\mathbf{r}}{dl}-3(\mathbf{m}_F\cdot \mathbf{r})(\mathbf{r}\cdot \frac{d\mathbf{r}}{dl})}{4\pi r^5}\\
&=\mu_0Q_m\left(\frac{\mathbf{r}_f}{4\pi r_f^3}-\frac{\mathbf{r}_i}{4\pi r_i^3}\right)\cdot\mathbf{m}_F=-\mathbf{m}_F\cdot \mathbf{B}_{str}
\end{split}
\end{equation}
where we have defined $Q_m=g_F\mu_BF_0/b$ with the spacing between atoms on the string $b\sim n_0^{-1/3}$, while $\mathbf{r}_i$ and $\mathbf{r}_f$ are the positions of the ends of the string relative to $\mathbf{m}_F$. This means that effectively the magnetic moment $\mathbf{m}_F$ only feels the magnetic field of the monopoles at the ends of the string, which carry monopole charges $\pm Q_m$ respectively. Remarkably, a closed Dirac string has no interaction energy with the other spins at all.

%It is then straightforward to derive the DDI energy between two open strings. We parameterize the first string as $\mathbf{r}(l)$ with ends located at $\mathbf{r}_i$ and $\mathbf{r}_f$, and denote the end points of the second string as $\mathbf{r}_i'$ and $\mathbf{r}_f'$. According to Eq. (\ref{EmF}), the total interaction energy between the two strings is simply given by

Further, it is straightforward to prove that the interaction energy between two Dirac strings is solely given by the interaction between the four monopoles at the ends of the strings. Consider two strings $L_1$ and $L_2$ as shown in Fig. \ref{string}(b), whose ends are located at $\mathbf{r}_i$, $\mathbf{r}_f$ and $\mathbf{r}_i'$, $\mathbf{r}_f'$ respectively. We shall parameterize string $L_1$ as $\mathbf{r}(l)$, and use the abbreviations $\mathbf{r}_i(l)=\mathbf{r}_i'-\mathbf{r}(l)$, $\mathbf{r}_f(l)=\mathbf{r}_f'-\mathbf{r}(l)$ to denote the displacement vectors from $\mathbf{r}(l)$ to the ends of string $L_2$. The magnetic moment of an infinitesimal segment $dl$ on string $L_1$ is then $\mathbf{m}_F=g_F\mu_BF_0 d\mathbf{r}(l)/b=Q_md\mathbf{r}(l)=-Q_md\mathbf{r}_i(l)=-Q_md\mathbf{r}_f(l)$. According to the result of Eq. (\ref{EmF}), the total DDI energy between the two strings is simply
\begin{equation}\label{EQ}
\begin{split}
&E_Q(L_1,L_2)=\mu_0Q_m^2\int d\mathbf{r}(l)\left[\frac{\mathbf{r}_f(l)}{4\pi r_f(l)^3}-\frac{\mathbf{r}_i(l)}{4\pi r_i(l)^3}\right]\\
&=\mu_0Q_m^2 \left[-\int_{\mathbf{r}_{if}}^{\mathbf{r}_{ff}}\frac{\mathbf{r}_f(l)d\mathbf{r}_f(l)}{4\pi r_f(l)^3}+\int_{\mathbf{r}_{ii}}^{\mathbf{r}_{fi}}\frac{\mathbf{r}_i(l)d\mathbf{r}_i(l)}{4\pi r_i(l)^3}\right]\\
&=\frac{\mu_0Q_m^2}{4\pi} \left(\frac{1}{r_{ff}}-\frac{1}{r_{if}}-\frac{1}{r_{fi}}+\frac{1}{r_{ii}}\right)\ ,
\end{split}
\end{equation}
where $r_{ff}$, $r_{if}$, $r_{fi}$ and $r_{ii}$ are the distances between monopoles at the ends of the two strings as shown in Fig. \ref{string}(b). The interaction energy between two monopoles is thus simply given by the Coulomb law. Accordingly, two parallel(anti-parallel) Dirac strings will repulse(attract) each other. Still, a closed string does not interact with any other strings. Therefore, the Dirac strings are almost free under DDI except for the monopole interactions between their ends.

In the continuum limit, we can view the flow lines of the local spin field $\bm{\mathcal{F}}(\mathbf{r})$ of the condensate as Dirac strings, each of which consists of a single column of atoms. We denote the cross sectional area of a string as $d\sigma$, which is around $n_0^{-2/3}$. The spacing between atoms on a string is then $b=(n_0d\sigma)^{-1}$, and the monopole charge can be expressed as $Q_m=g_F\mu_Bn_0F_0d\sigma$, which is proportional to the cross sectional area $d\sigma$.

So far we have seen the bulk of a Dirac string has no interaction with other strings at all under DDI. However, as we shall shown in the below, a Dirac string has a negative self-interaction bulk energy coming from the DDI.
%Though the bulk of a Dirac string has no interaction with the other strings at all, it has a negative self interaction energy.
To understand the origin of the self-interaction energy, consider a straight Dirac string of length $L$. Since all the spins are aligned head-to-tail, there is an attractive DDI energy $-2(\lambda/n_0^2)(1/4\pi r^{3})$ between any two of them, where $r$ is the distance between them. This gives rise to the self-interaction energy of the string. In the limit $b<<L$, the attraction energy felt by a spin on the string is approximately
\[E_{F_0}\approx-\frac{\lambda}{n_0^2}\sum_{N=1}^{\infty}\frac{1}{\pi}\left(\frac{1}{Nb}\right)^3\propto-\frac{\lambda}{n_0}\ ,\]
where we have used the fact $1/b^3\propto n_0$. Thus, the self-attraction energy $E_{F_0}$ of each spin is a constant not sensitive to the string length $L$. %The self-interaction energy of the string is then $E_{str}^{(d)}(L)=(1/2)E_{F_0}(L/b)=(1/2)E_{F_0}n_0Ld\sigma$,

%Consider a spin on a Dirac string of length $L\gg b$. Locally, we can view the string as a straight line, so the spin is attracted by the other spins on the line. The attraction energy the spin feels is then estimated to be $-E_{F_0}\propto \mu_0(g_F\mu_BF_0)^2\sum_{N=1}^{\infty}(1/Nb)^3\propto\mu_0(g_F\mu_BF_0)^2n_0=\lambda/n_0$.
The above estimation, however, cannot give the coefficient in front of $-\lambda/n_0$ in the expression of $E_{F_0}$, since we do not know the exact relation between $1/b^3$ and $n_0$ in the continuum limit. Instead, we can find out the coefficient by directly computing the DDI energy of a simple concrete example of spin texture. Consider a 3D spherical condensate with all spins polarized along $z$ direction as shown in Fig. \ref{sphere}(a). The total DDI energy $E_d$ felt by the spin at the center can be directly calculated in spherical coordinates from its definition:
\begin{equation}\label{ED1}
E_d=\frac{\lambda}{n_0}\int_0^R 2\pi r^2dr\int_0^\pi d\theta \sin\theta \frac{1-3\cos^2\theta}{4\pi r^3}=0\ ,
\end{equation}
where $R$ is the trap radius, and $\theta$ is the polar angle. On the other hand, in the picture of Dirac strings, the DDI energy $E_d$ felt by the spin consists of the attraction energy $E_{F_0}$ from its own string and the interaction with monopoles at the ends of all the strings derived in Eq. (\ref{EmF}). In cylindrical coordinates $(\rho,\phi,z)$, the cross sectional area of a string is $d\sigma=\rho d\rho d\phi$, and this energy is calculated as
\begin{equation}\label{ED2}
E_d=E_{F_0}+\frac{2\lambda}{n_0}\int_0^R 2\pi\rho d\rho\frac{\sqrt{R^2-\rho^2}}{4\pi R^3}=E_{F_0}+\frac{\lambda}{3n_0}\ .
\end{equation}
By equalizing Eq. (\ref{ED1}) and Eq. (\ref{ED2}), we see the attraction energy
\[E_{F_0}=-\frac{\lambda}{3n_0}\ .\]
The total self-interaction energy of a Dirac string of length $L$ and of cross sectional area $d\sigma$ is then
\begin{equation}
E_{str}^{(d)}(L)=\frac{E_{F_0}}{2}\times n_0Ld\sigma=-\frac{1}{6}\lambda L d\sigma\ .
\end{equation}
This indicates that a Dirac string has a negative string tension $T_{str}=-\lambda d\sigma/6$, and therefore the whole space will prefer to be filled with Dirac strings. The action for a single Dirac string can be written as $\mathcal{S}_{str}=(T_{str}/\hbar)\int dtdl=T_{str}\Sigma/\hbar$, where $\Sigma$ is the worldsheet area of the string. The Dirac string therefore corresponds to a classical bosonic string theory with a U($1$) gauge group \cite{Green1987}.

Now we turn to the kinetic energy $H_0$ in Eq. (\ref{HMF}), and interpret it in the language of Dirac strings. By writing the gradient operator into $\nabla=\nabla_\parallel+\nabla_\perp$, where $\parallel$ and $\perp$ stands for parallel and perpendicular to the string respectively, we can divide $H_0$ into the following two parts: the first parallel part contributes an additional term to the string self energy. This is seen by noticing that $\nabla_\parallel\bm{\mathcal{F}}=d \bm{\mathcal{F}}/dl=d^2\mathbf{r}/dl^2$, so the string self energy now reads
\begin{equation}\label{Estr}
E_{str}(L)=d\sigma\int_0^L dl\left[-\frac{1}{6}\lambda+\frac{\alpha}{M}\left(\frac{d^2\mathbf{r}}{dl^2}\right)^2\right]\ .
\end{equation}
Note that $|d^2\mathbf{r}/dl^2|$ is the curvature of the string. The second perpendicular part induces a nonnegative contact interaction between two Dirac strings $L_1$ and $L_2$ (see Appendix \ref{StrInt}):
\begin{equation}\label{EI}
\begin{split}
&E_{I}(L_1,L_2)=d\sigma_1 d\sigma_2 \frac{\alpha}{M}\int_0^{L_1} dl_1\int_0^{L_2} dl_2\\
&\quad\times\left(1-\frac{d\mathbf{r}_1}{dl_1}\cdot \frac{d\mathbf{r}_2}{dl_2}\right)\nabla_{\perp}^2\delta(\mathbf{r}_{1}-\mathbf{r}_{2})\ ,
\end{split}
\end{equation}
where $\nabla^2_\perp$ is defined as $\nabla^2_\perp=\nabla^2-\partial^2_{l_1}$, and $\mathbf{r}_1=\mathbf{r}_1(l_1)$, $\mathbf{r}_2=\mathbf{r}_2(l_2)$. %where we have used the notations $\mathbf{r}_{12}=\mathbf{r}_1(l_1)-\mathbf{r}_2(l_2)$ and $\bm{\mathcal{F}}_i=d\mathbf{r}_i(l_i)/dl_i$.
Though looks complicated, this contact interaction simply implies that two nearby Dirac strings prefer to be coplanar, namely, $|d\mathbf{r}_1/dl_1-d\mathbf{r}_2/dl_2|/|\mathbf{r}_1-\mathbf{r}_2|\rightarrow0$ as $|\mathbf{r}_1-\mathbf{r}_2|\rightarrow 0$.

Therefore, we have seen the ferromagnetic dipolar condensate at the mean-field level is dual to a Dirac string gas with a string self energy $E_{str}$, a monopole interaction energy $E_Q$ and a contact interaction energy $E_{I}$, given in Eq. (\ref{Estr}), Eq. (\ref{EQ}) and Eq. (\ref{EI}), respectively. With $E_{str}$ mainly coming from the DDI $H_D$ and $E_I$ coming from $H_0$, such a duality between individual spins and Dirac strings interchanges the roles of the kinetic energy and the interaction energy in some sense, and is thus a strong-weak duality.

\section{Application to spherical trap geometries}
The ground state spin texture can be more easily derived in the Dirac string gas picture. First, since an open string can always lower its monopole interaction energy by attaching its two ends together and forming a closed string, a strong dipolar condensate will prefer as many closed strings as possible. Further, to reduce the kinetic energy $H_0$, the close strings will tend to have a smaller average curvature and be locally coplanar to each other. These facts constitute the guiding rules for identifying the spin textures in a condensate under various trap geometries.

\begin{figure}
\includegraphics[width=3.2in]{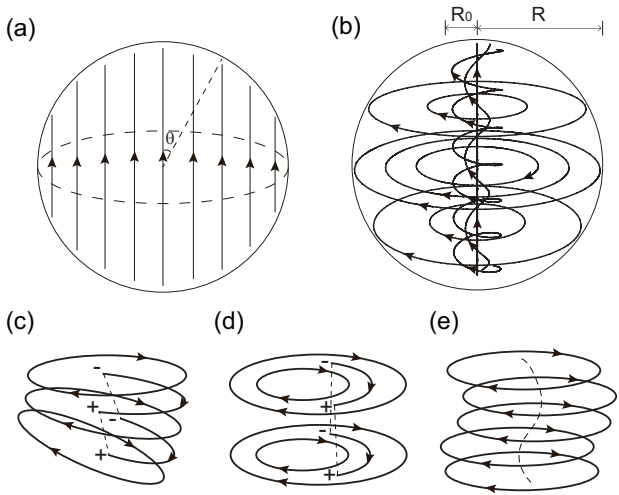}
\caption{(a) In the absence of dipolar interaction, all the spins will be polarized along the same direction. (b) The meron-like vortex spin texture of spherical dipolar condensates. (c), (d) and (e) shows several perturbation patterns that may arise in the spin texture, all of which increase the total energy. \label{sphere}}
\end{figure}

%\emph{meron-like spin texture in 3D, perturbation to the texture}
We first consider a dipolar condensate in a 3D spherical square well trap potential of radius $R$. Based on the guiding rules above, the most natural spin texture one can expect is a meron-like vortex texture as shown in Fig. \ref{sphere}(b). Since circular closed strings possess the smallest average curvature, the region $\rho>R_0$ (in cylindrical coordinates) prefers to be filled with circular strings parallel to each other, where $R_0$ is a characteristic length scale to be determined. All the circular strings are symmetric about the $z$ axis, so that locally they are always coplanar. At $\rho<R_0$, the circular strings will have too large curvatures, and thus will gradually give way to spiral strings proceeding along the $z$ direction to lower the kinetic energy $H_0$. Finally, the string becomes vertical and straight at the central axis $\rho=0$. Consequently, there will be monopoles distributed on the $\rho<R_0$ regions of the sphere's surface, with a monopole area density $n_A\approx \pm Q_m/d\sigma=\pm g_F\mu_Bn_0F_0$. The total energy of the dipolar condensate can be estimated as
\begin{equation}\label{E3D}
\begin{split}
&E_{3D}\approx -\frac{\lambda}{6}V+\int_{\rho\ge R_0}^R d^3\mathbf{r}\frac{\alpha}{M}\frac{1}{\rho^2}+\frac{2\mu_0(\pi R_0^2n_A)^2}{4\pi R_0} \\
&\approx -\frac{\lambda}{6}V+\frac{4\pi\alpha}{M}R\log\frac{R}{R_0}+\frac{\pi}{2}\lambda R_0^3\ ,
\end{split}
\end{equation}
where $V=4\pi R^3/3$ is the volume of the condensate, and $R_0/R$ is assumed small. The second term is the kinetic energy $H_0$ mainly coming from the curvatures of the strings, while the third term is the monopole interaction energy estimated by taking $R_0$ as the average distance between monopoles \cite{attraction3D}. By minimizing the total energy with respect to $R_0$, we get the characteristic length
\[R_0=(8\alpha /3\lambda M)^{1/3}R^{1/3}\ .\]
The condition for the meron-like spin texture to arise is then roughly $R>R_0$, or $R>(8\alpha/3\lambda M)^{1/2}$. Therefore, the spin textures are more likely to arise for massive and large spin atoms.

\begin{figure}
\includegraphics[width=3.3in]{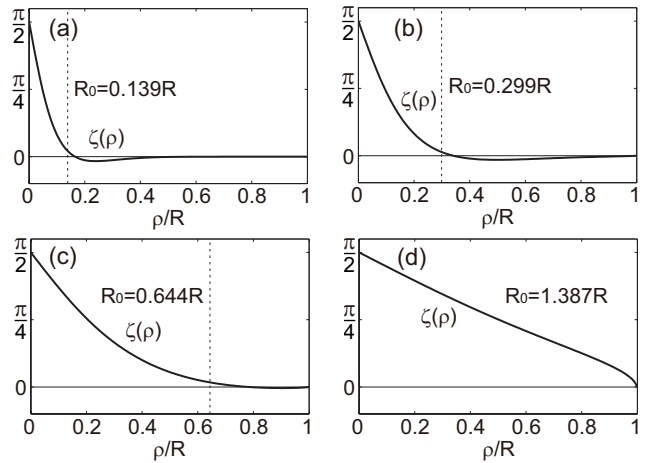}
\caption{The ground state angle function $\zeta(\rho)$ calculated for different values of $R_0=(8\alpha /3\lambda M)^{1/3}R^{1/3}$. The dashed lines denote the positions of $R_0$. When $R_0<R$, $R_0$ is a rather precise boundary separating the $\zeta(\rho)\neq0$ region and the $\zeta(\rho)\approx0$ region. \label{angle}}
\end{figure}

It is easy to see the meron-like spin texture is energetically stable. When perturbed, the circular strings may be tilted, misaligned, or distorted. In any case, the total energy is raised higher. In the examples shown in Fig. \ref{sphere}(c) and Fig. \ref{sphere}(d), parallel open strings have to arise to keep the strings dense, which then cost an extra monopole interaction energy. In another case shown in Fig. \ref{sphere} (e), no monopole occurs, but the strings are no longer locally coplanar to each other, which also increases the kinetic energy $H_0$. These perturbation patterns may have interesting dynamics, which is open to studies in the future.

To verify the precision of our estimation above, we perform a numerical calculation based on the following spin texture ansatz, expressed in cylindrical coordinates ($\rho,\phi,z$) as:
\begin{equation}
\bm{\mathcal{F}}(\mathbf{r})=\mathbf{e}_\phi\cos\zeta(\rho)+\mathbf{e}_z\sin\zeta(\rho)\ ,
\end{equation}
where $\zeta(\rho)$ is a function of $\rho$ only. This ansatz simply describes the spin texture shown in Fig. \ref{sphere}(b). $\zeta(\rho)$ has the physical meaning of the proceeding angle of a spiral string at radius $\rho$, and the string becomes circular when $\zeta(\rho)=0$.  Fig. \ref{angle} shows $\zeta(\rho)$ calculated numerically for various values of $R_0$, and the details of the calculations can be found in Appendix \ref{Numerical}. As is seen, our estimation of $R_0$ gives a rather precise boundary between the region of spiral strings ($\zeta(\rho)\neq0$) and the region of circular strings ($\zeta(\rho)\approx0$), even for $R_0/R$ that is no longer very small.

\section{Applications to cigar and pancake trap geometries}
\begin{figure}
\includegraphics[width=3.2in]{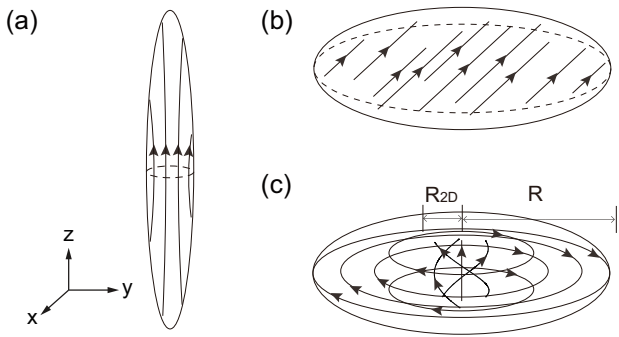}
\caption{(a) The flare texture in a quasi-1D cigar trap. (b) The in-plane polarized texture in a quasi-2D pancake trap. (c) The meron texture in the pancake trap. \label{trap}}
\end{figure}

This analysis can be readily applied to the dipolar condensates in other trap geometries. A practical and important geometry is the axially symmetric square well trap where the boundary is given by $x^2+y^2+(z/A)^2=R^2$. By setting $A\gg1$, we get a quasi-1D cigar trap of length $L_{c}=2AR$. In the extremely quasi-1D case, closed strings are no longer energetically favorable, since their curvatures are always too large. Instead, vertical strings along $z$ direction are preferred, so that the amount of monopoles on the surface is minimal and the monopole energy is the smallest. In a realistic trap, the particle number density $n_0$ is not constant and usually lower near the surface. To validate our theoretical formulation in Sec. \ref{theory}, we need to keep $b=(n_0d\sigma)^{-1}$ constant for a string, so the effective cross sectional area density of strings $n_s\sim1/d\sigma\propto n_0$ is also lower near the surface. In the cigar trap case, the vertical strings are then forced to be bent outward to have a lower density near the surface, forming a flare texture as shown in Fig. \ref{trap}(a). By an energy estimation similar to what we did for spherical traps, one can show that circular closed strings begin to occur when $A\sim (\lambda M/\alpha)^{1/3}L_c^{2/3}$ or smaller, leading to a crossover from the flare texture to the 3D meron-like vortex texture.

In the opposite limit $A\ll1$, the trap is a quasi-2D pancake of radius $R$. Correspondingly, there are two candidate ground state spin textures. The first spin texture is shown in Fig. \ref{trap}(b), where all strings are in-plane and polarized along the same direction to minimize the amount of surface monopoles. The total energy of this texture consists of the string self-energy and the surface monopole interaction energy, or explicitly
\begin{equation}
E_{2D}^{(I)}\approx-\frac{\lambda}{6}V+w\lambda R^3A^2\log \frac{1}{A} \ ,
\end{equation}
where $V$ is the volume of the condensate, and $w>0$ is a numerical factor. The second spin texture is the 2D meron texture shown in Fig. \ref{trap}(c). Similar to the 3D case, there is a characteristic radius $R_{2D}$. The region outside $R_{2D}$ is filled with in-plane circular strings, while inside $R_{2D}$ strings become spiral and finally vertical at the center. In the case the half height of the pancake $z_h=AR\ll R_{2D}$, the total energy of the configuration can be estimated as \cite{monopole2D}
\begin{equation}
E_{2D}^{(II)}\approx-\frac{\lambda}{6}V+\frac{4\pi\alpha}{M}z_h\log\frac{R}{R_{2D}}+\frac{\pi\lambda}{2} z_hR_{2D}^2\ .
\end{equation}
Minimizing the energy we get $R_{2D}=(4\alpha/\lambda M)^{1/2}$, which is independent of $z_h$ or $R$. At a fixed half height $z_h$, one can show that the in-plane polarized texture is favored when $R<R_c$, where $R_c$ is a critical radius determined by solving $E_{2D}^{(I)}=E_{2D}^{(II)}$, while the meron texture arises when $R>R_c$. We note the two textures are separated by a phase transition at $R=R_c$ instead of a crossover.

\begin{figure}
\includegraphics[width=3.2in]{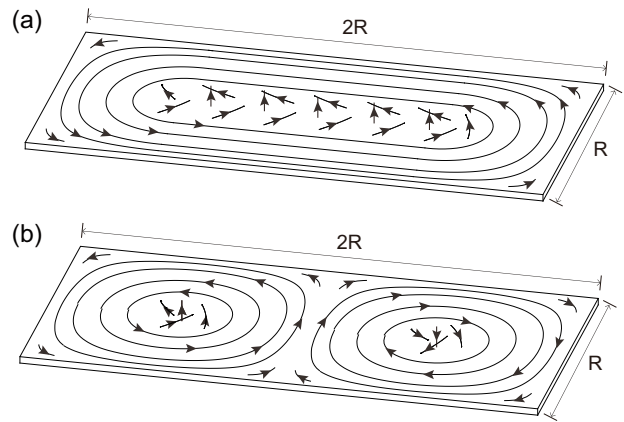}
\caption{In a large enough quasi-2D trap with an aspect ratio $q=2$, the single meron in (a) costs more kinetic energy than the two antiparallel merons in (b).\label{ratio}}
\end{figure}

%\emph{multiple meron texture in 2D with non-square aspect ratios}
When the pancake trap is not axially symmetric, multiple merons may occur. As an example, in a rectangular trap with aspect ratio $q=2$, the single-meron texture shown in Fig. \ref{ratio}(a) costs a kinetic energy $\propto z_hR/R_{2D}$ near the center. Instead, the two-meron texture in Fig. \ref{ratio}(b) only costs a kinetic energy $\propto z_h\log(R/R_{2D})$, and thus will be favored at large $R$. The central spins of the two merons are aligned antiparallel to each other to further lower the DDI energy. Similarly, one may expect a $q$-meron spin texture to arise for a trap with aspect ratio $q$, which agrees well with the numerical results obtained in Ref. \cite{Zhang2010}.

\section{Conclusion}
In conclusion, we have shown that the duality to the Dirac string gas picture is an efficient semi-analytical way of determining the spontaneous spin texture in a dipolar spinor condensate. The total dipolar energy can be conveniently estimated in the perspective of Dirac strings. Closed Dirac strings with curvatures as small as possible are generically preferred in the ground state. We expect this method to be further employed in future work to study the spin textures in external fields and the dynamics of spin textures. In addition, it will also be useful and intriguing to examine the possibility of constructing a quantum version of this duality as a generalization of the mean-field level duality presented here.

\section{Acknowledgements}
The author thanks Prof. S. C. Zhang for carefully reading the manuscript, and the referees for their valuable suggestions.
%The author is supported by the Stanford Graduate Fellow program in Science and Engineering under the William R. and Sara Hart Kimball Fellowship.

\appendix

\section{Mean field description of ferromagnetic spinor condensates}\label{MFdescription}

Spin $F$ bosons with DDI are most generally described by a Hamiltonian $\mathcal{H}=\mathcal{H}_0+\mathcal{H}_D$ , where $\mathcal{H}_0$ represents the kinetic energy and the local interaction \cite{Lian2012}, and $\mathcal{H}_D$ is the long range DDI. In terms of the boson annihilation (creation) operators $\psi=(\psi_F,\cdots,\psi_{-F})^T$, $\mathcal{H}_0$ is given by
\begin{equation}
\mathcal{H}_0=\int d^3\mathbf{r}\left[\psi^\dag\left(-\frac{\hbar^2\nabla^2}{2M}-\mu\right)\psi+\sum_{J=0}^{F}g^I_{2J} \hat{P}_{2J}\right]\ ,
\end{equation}
where $\hat{P}_{2J}=\sum_m(\psi^\dag\otimes\psi^\dag)|2J,m\rangle\langle2J,m|(\psi\otimes\psi)$ is the local two-particle interaction projected into the total spin $2J$ channel, and $g^I_{2J}$ is the corresponding interaction strength, and $g^I_{2J}=2\pi\hbar^2 a_{2J}/M$ in terms of the s-wave scattering length $a_{2J}$. The whole Hamiltonian is $U(1)\times SU(2)$ symmetric. In the mean-field approximation, $\langle\psi_m(\mathbf{r})\rangle=\Psi_m(\mathbf{r})$, where $\Psi_m(\mathbf{r})=\sqrt{n(\mathbf{r})}\varphi_m(\mathbf{r})$ is the spinor order parameter. The mean field energy of $\mathcal{H}_0$ can then be written in the form
\begin{equation}\label{MFenergy}
\mathcal{H}_0=\int d^3\mathbf{r}\left[\frac{\hbar^2}{2M}|\nabla \sqrt{n(\mathbf{r})}\varphi(\mathbf{r})|^2+ V(n(\mathbf{r}),\varphi(\mathbf{r}))\right]\ ,
\end{equation}
where $V(n,\varphi)$ is invariant under $U(1)\times SU(2)$ transformations of $\varphi$. For a ferromagnetic spinor condensate, both the U(1) and SU(2) symmetries are spontaneously broken, and there are two Goldstone modes corresponding to the U(1) phase fluctuation and the spin direction fluctuation respectively \cite{Lian2012}. At low energies, by integrating out the quantum fluctuation in density $n(\mathbf{r})$ in the path integral formalism, we can reduce the mean-field energy (\ref{MFenergy}) to
\begin{equation}
\begin{split}
&H_0=\int d^3\mathbf{r}\left[\frac{\hbar^2(\partial_t\phi_{SF})^2}{4(\partial^2V/\partial n^2)}+ \frac{\hbar^2n_0}{2M}|\nabla\varphi|^2\right]\\
=&\int d^3\mathbf{r}\left[\frac{\hbar^2(\partial_t\phi_{SF})^2}{4(\partial^2V/\partial n^2)}+ \frac{\hbar^2n_0}{2M}(\nabla\phi_{SF})^2+\frac{\alpha}{M} (\nabla\bm{\mathcal{F}})^2\right]\ ,
\end{split}
\end{equation}
where $\phi_{SF}(\mathbf{r})$ is the U(1) phase, $\bm{\mathcal{F}}(\mathbf{r})=\varphi^\dag\mathbf{F}\varphi/F_0$ is the normalized local spin, and the coefficient can be determined to be $\alpha=n_0\hbar^2[F(F+1)-F_0^2]/4$. We note that the local interactions $g^I_{2J}$ only contribute to the parameter $\partial^2V/\partial n^2$ in the first time dependent term, and do not enter the expression of the coefficient $\alpha$.

Similarly, the mean field DDI energy $H_D$ in Eq. (\ref{HMF}) can be deduced by substituting the operator $\psi_m$ with the mean field value $\Psi_m=\sqrt{n_0}\varphi_m$.

Since the DDI only depends on the spin field $\bm{\mathcal{F}}(\mathbf{r})$, the ground state always tends to have a constant U(1) phase $\phi_{SF}$ provided the spin field $\bm{\mathcal{F}}(\mathbf{r})$ is smooth enough. Therefore, the effective low energy Hamiltonian of the system reduces to Eq. (\ref{HMF}).

\section{The kinetic energy $H_0$ rewritten in the Dirac string picture}\label{StrInt}
To interpret the kinetic energy $H_0$ in the Dirac string picture, we first note that it can be rewritten as
\begin{equation}
\begin{split}
&H_0=-\frac{\alpha}{M}\int d^3\mathbf{r}\ \bm{\mathcal{F}}(\mathbf{r})\cdot\nabla^2\bm{\mathcal{F}}(\mathbf{r})\\
&=-\frac{\alpha}{M}\int d^3\mathbf{r}_1d^3\mathbf{r}_2\delta(\mathbf{r}_1-\mathbf{r}_2)\bm{\mathcal{F}}(\mathbf{r}_2)\cdot\nabla_1^2\bm{\mathcal{F}}(\mathbf{r}_1)\ ,
\end{split}
\end{equation}
where $\nabla_1^2$ is the Laplace operator with respect to $\mathbf{r}_1$. In the Dirac string picture, we can write the volume element $d^3\mathbf{r}_i$ as $dl_id\sigma_i$, and we have $\bm{\mathcal{F}}(\mathbf{r}_i)=d\mathbf{r}_i/dl_i$, for $i=1,2$. The kinetic energy then becomes
\begin{equation}
H_0=-\frac{\alpha}{M}\int d\sigma_1d\sigma_2dl_1dl_2\delta(\mathbf{r}_1-\mathbf{r}_2)\frac{d\mathbf{r}_2}{dl_2}\cdot\nabla_1^2\frac{d\mathbf{r}_1}{dl_1}\ .
\end{equation}
By partial integrating the expression two times, we find
\begin{equation}
\begin{split}
&H_0=-\frac{\alpha}{M}\int d\sigma_1d\sigma_2dl_1dl_2\frac{d\mathbf{r}_2}{dl_2}\cdot\frac{d\mathbf{r}_1}{dl_1}\nabla^2_1\delta(\mathbf{r}_1-\mathbf{r}_2)\\
&=\frac{\alpha}{M}\int d\sigma_1d\sigma_2dl_1dl_2\left[1-\frac{d\mathbf{r}_2}{dl_2}\cdot\frac{d\mathbf{r}_1}{dl_1}\right]\\
&\qquad\times\left[\nabla^2_\perp\delta(\mathbf{r}_1-\mathbf{r}_2)+\partial_{l_1}^2\delta(\mathbf{r}_1-\mathbf{r}_2)\right]\\
&=\frac{\alpha}{M}\int d\sigma_1d\sigma_2dl_1dl_2\left[1-\frac{d\mathbf{r}_2}{dl_2}\cdot\frac{d\mathbf{r}_1}{dl_1}\right]\\
&\quad\times\nabla^2_\perp\delta(\mathbf{r}_1-\mathbf{r}_2)-\frac{\alpha}{M}\int d\sigma_1dl_1\ \frac{d\mathbf{r}_1}{dl_1}\cdot\frac{d^3\mathbf{r}_1}{dl_1^3}\\
&=\frac{\alpha}{M}\int d\sigma_1d\sigma_2dl_1dl_2\left[1-\frac{d\mathbf{r}_2}{dl_2}\cdot\frac{d\mathbf{r}_1}{dl_1}\right]\\
&\quad\times\nabla^2_\perp\delta(\mathbf{r}_1-\mathbf{r}_2)+\frac{\alpha}{M}\int d\sigma dl\ \left(\frac{d^2\mathbf{r}}{dl^2}\right)^2\ ,\\
\end{split}
\end{equation}
where $\nabla_\perp^2=\nabla^2-\partial_{l_1}^2$. As is seen, the first term is a nonnegative contact interaction between two Dirac strings, while the second term is an additional contribution to the self-energy of a Dirac string.

\section{Numerical calculations of spin texture in spherical traps}\label{Numerical}
In Eq. (\ref{E3D}) we have derived an estimation of the total energy of the spin texture in a spherical trap from the Dirac string perspective, and show that the meron texture has a core size $R_0=(8\alpha /3\lambda M)^{1/3}R^{1/3}$. Here we present a more accurate numerical calculation of the configuration of the meron texture. Based on the reasoning given in the main text, the following spin texture ansatz is appropriate, written in spherical coordinates ($\rho,\phi,z$) as:
\begin{equation}
\bm{\mathcal{F}}(\mathbf{r})=\mathbf{e}_\phi\cos\zeta(\rho)+\mathbf{e}_z\sin\zeta(\rho)\ ,
\end{equation}
bounded in the square well trap $\rho^2+z^2\le R^2$. As is shown in Fig. \ref{sphere}(b), the monopole charges of this ansatz are distributed on the surface of the trap, with an amount $\pm n_A\rho d\rho d\phi\sin\zeta(\rho)$ in the projected area $\rho d\rho d\phi$, where we have defined $n_A=g_F\mu_B n_0F_0$. The total energy of the spin texture is then the summation of the string self energy, the kinetic energy and the monopole energy:
\begin{equation}
\begin{split}
&E_{3D}=-\frac{\lambda}{6}V+\int d^3\mathbf{r}\frac{\alpha}{M}(\nabla\bm{\mathcal{F}})^2+E_Q\\
&=-\frac{\lambda}{6}V+\int_0^R d\rho \rho\sqrt{R^2-\rho^2}\frac{4\pi\alpha}{M}\left[\left(\frac{d\zeta}{d\rho}\right)^2+\frac{\cos^2\zeta}{\rho^2}\right]\\
&+\int_0^R d\rho d\rho' \int_0^{2\pi}d\phi\frac{\lambda\rho\rho'\sin\zeta\sin\zeta'}{2\sqrt{2}}\left(\frac{1}{D_+}-\frac{1}{D_-}\right)\ ,
\end{split}
\end{equation}
where
\begin{equation}
D_\pm=\sqrt{R^2-\rho\rho'\cos\phi\mp\sqrt{(R^2-\rho^2)(R^2-\rho'^2)}}\ ,
\end{equation}
and we have used the abbreviations $\zeta=\zeta(\rho)$, $\zeta'=\zeta(\rho')$. $\sqrt{2}D_\pm$ has the physical meaning of the distance between two monopole charges ($+$ for charges on the same hemisphere, $-$ for charges on the opposite hemisphere). The integration $\int d\phi/D_\pm$ can be expressed in terms of the elliptic K function to further simplify the energy functional.

This energy functional of $\zeta(\rho)$ can be minimized via the steepest descent method. In practice, we start with an arbitrary function of $\zeta(\rho)$, and add to the function an increment
\[\delta\zeta(\rho)=-\epsilon\nabla_{\zeta(\rho)}E_{3D}\ ,\]
repeatedly until the process converges, where $\epsilon$ is a small number. Provided $\epsilon$ is small enough, the numerical calculation converges well. The resulting functions $\zeta(\rho)$ for various values of $R_0$ are shown in Fig. \ref{angle}.

\bibliography{Dipole-ref}

\begin{thebibliography}{19}
\bibitem [{att()}]{attraction3D}%
  \BibitemOpen
  \href@noop {} {}\bibinfo {note} {Here the attraction between monopoles on the
  north pole and those on the south pole is neglected, which is around
  $-\lambda R_0^4/R$ and therefore smaller by a factor $R_0/R$.}\BibitemShut
  {Stop}%
\bibitem [{mon()}]{monopole2D}%
  \BibitemOpen
  \href@noop {} {}\bibinfo {note} {More precisely, the monopole energy (the
  third term of $E_{2D}^{(II)}$) is
  $(\pi\lambda/2)R^2_{2D}(R_{2D}+z_h-\sqrt{R_{2D}^2+z_h^2})$.}\BibitemShut
  {Stop}%
\end{thebibliography}

\end{document}